\documentclass[twocolumn,english]{revtex4}
\usepackage[T1]{fontenc}
\usepackage[latin9]{inputenc}
\usepackage{babel}
\usepackage{float}
\usepackage{amsmath}
\usepackage{amssymb}
\usepackage{graphicx}
\usepackage[unicode=true,pdfusetitle,
bookmarks=true,bookmarksnumbered=false,bookmarksopen=false,
breaklinks=false,pdfborder={0 0 0},backref=false,colorlinks=false]
{hyperref}

\makeatletter
\@ifundefined{textcolor}{}
{%
\definecolor{BLACK}{gray}{0}
\definecolor{WHITE}{gray}{1}
\definecolor{RED}{rgb}{1,0,0}
\definecolor{GREEN}{rgb}{0,1,0}
\definecolor{BLUE}{rgb}{0,0,1}
\definecolor{CYAN}{cmyk}{1,0,0,0}
\definecolor{MAGENTA}{cmyk}{0,1,0,0}
\definecolor{YELLOW}{cmyk}{0,0,1,0}
}

\@ifundefined{definecolor}
{\usepackage{color}}{}

\makeatother

\begin{document}

\title{Classical simulation of entangled states}

\author{H. M. Bharath$^{1}$ }

\email{bharath.hm@gatech.edu}

\selectlanguage{english}%

\author{V. Ravishankar$^{2,3}$}
\email{vravi@iitk.ac.in, vravi@physics.iitd.ac.in}

\selectlanguage{english}%

\affiliation{$^{1}$ School of Physics, Georgia Institute of Technology, Atlanta, Georgia-30332, (U.S.A)}
\affiliation{$^{2}$Department of Physics, Indian Institute of Technology, Kanpur - 208016 (India)}
\affiliation{$^{3}$Department of Physics, Indian Institute of Technology, Delhi -110016 (India)}
\begin{abstract}
Description of nonclassicality of states has hitherto been through violation of Bell inequality and non-separability, with the latter being a stronger constraint. In this paper, we show that this can be further sharpened, by introducing the concept of classical simulation. A state admits classical simulation if it can be mimicked fully by a separable state of higher dimension. A nonclassical state, which we call exceptional, does not admit classical simulation. Focusing on two qubit states, we show that exceptionality is more stringent than violation of Bell inequality, and involves an intricate interplay of coherence and entanglement. The new criterion is shown to provide a natural description of entangled states which respect Bell inequality, and also a way of enumerating the classical resources that are required to simulate a quantum state. Possible implications to quantum dynamics and quantum information are briefly touched upon.

\end{abstract}
\maketitle

\section {Introduction}
The problem of distinguishing the truly quantum properties of a system from those which can be realized in classical systems has received close attention ever since the birth of quantum mechanics \cite{sch, epr,bohm, bell,chsh, ch,werner}. Two key notions that dominate this discourse are nonlocality \cite{sch,epr,bohm,bell} and non classical correlations \cite{werner}. Nonlocality is expressed in terms of violation of Bell-type inequalities, while entanglement may be described in terms of non-separability -- which implies the existence of nonclassical correlations. However, one knows that there are states which exhibit quantum correlations and are therefore entangled, but nevertheless admit a local hidden variable (LHV) description \cite{werner}. This leaves open the possibility that there could be other formulations of quantumness of a system which combine - and refine - the aspects of non-separability and nonlocality.

The purpose of this paper is to revisit and reformulate the problem of characterizing the quantumness of a system by introducing the notion of classical simulation. It is necessary to be clear on what we mean by classical simulation since it is used in more than one sense in literature. A precise definition will be given in Sect.3. Suffice to mention here that by simulability we mean the existence of higher dimensional separable states which can mimic a lower dimensional entangled state. This definition is quite close to separability or locality, but weaker than either of them.
Consequently, the criterion for quantumness that follows -- which we call exceptionality -- is more stringent.
In particular, violation of bell inequality is not sufficient for a state to defy classical simulation.

In order to keep things simple, we focus on the two qubit case, and present the results for three representative classes of entangled states. We employ the language of spin for convenience. The
proofs are fairly straight forward. Nevertheless, for the sake of uncluttered formulation and to keep the paper self contained, we start with a few preliminaries. They also serve to motivate the problem.

\section{Preliminaries} We start with a recapitulation of two key concepts, non-separability and spin coherent states. The former is central to the paper, and the latter plays a pivotal role in motivating the definition of classical simulation.

\subsection{Non separability}
Let states of two quantum systems $A, B$ be described in Hilbert spaces ${\cal H}^M, {\cal H}^N$ of respective dimensions $M,N$. States of the composite system $AB$ are then described in ${\cal H}^{MN} \equiv {\cal H}^M \otimes {\cal H}^N$. Recall that a state
$\rho^{AB}$ is defined to be separable if it admits a resolution
\begin{equation}
\rho^{AB}=\sum_{i}p_{i}\rho_{i}^{A}\otimes\rho_{i}^{B}
\end{equation}
where the weights $p_i \ge 0$ and the states $\rho_i^{A}~ (\rho_i^B) \in {\cal H}^{M} ~({\cal H}^N)$ \cite{werner}. In particular, if $\rho^{AB}$ is pure, the summation collapses to a single term. In the parlance of quantum correlations, separable states are deemed to be classical because
the expansion in RHS in Eq.1 expresses essentially a universal characteristic of all classical bipartite states.
Indeed, if $p(a, \alpha)$, $ 1\le a \le M;~ 1\le \alpha \le N$, be  the joint probability for two events $A_i, B_{\alpha}$,   it can always be expanded as
\begin{equation}
p(a, \alpha)= \sum_{b, \beta}p(b, \beta)\delta_{b, a}\delta_{\beta, \alpha}
\end{equation}
where the Kronecker deltas have the significance of "pure states" of respective subsystems, and $p(b, \beta)$ is the weight associated with the joint occurrence the product state $ \delta_{b, a}\delta_{\beta, \alpha}$. Thus, the notion of non-separability \cite{werner} implies that there are quantum states which do not share this property. Furthermore,
correlations in separable states have the form
\begin{eqnarray}
\langle O_1^A O_2^B\rangle _{\rho^{AB}} & = & \sum_{i} p_i Tr \{ O_1^A \rho_i^A \}Tr \{ O_2^B \rho_i^B \} \nonumber \\
& \equiv & \sum_i p_i\langle O_1^A \rangle_{\rho_i^A}\langle O_2^B \rangle_{\rho_i^B}
\end{eqnarray}
showing that the correlation is entirely attributable to the weights $p_i$ which, indeed, play the role of LHV. The converse is, of course, not true; a non-separable state may still admit a LHV description, as may be verified by showing that they respect Bell like inequalities \cite{werner}.

\subsection{Spin coherent states (SCS)} Coherent states
are well known in the context of classical quantum correspondence, and are employed in almost all branches of Physics. See, e.g., \cite{sudarshan,arechhi, perel,zhang,scully}. For our purposes, it is sufficient to consider spin coherent states (SCS), also known as atomic coherent states.
Recall that SCS for a spin $S$ particle may be generated by the action of the rotation group (more precisely, $SU(2)$) on
the state with the highest weight, $|S_z= +S\rangle$ \footnote{Some authors employ the state $|-S \rangle$ instead as the fiducial state.}:
\begin{equation}
\vert \hat{n}(\theta,\phi) \rangle \equiv e^{-iS_{z}\phi}e^{-iS_{y}\theta}e^{-iS_{z}\psi}|S\rangle
\end{equation}
in writing which the unphysical overall phase introduced by $e^{-iS_{z}\psi}$ is dropped. They satisfy the relations
\begin{eqnarray}
\langle \hat{n}(\theta,\phi)\vert \hat{S} \vert \hat{n}(\theta,\phi)\rangle & = & \hat{n}(\theta,\phi) \nonumber \\
\vert\langle\hat{n}|\hat{n'}\rangle \vert^2 & = & \Big(\frac{1+\hat{n}\cdot \hat{n'}}{2}\Big)^{2S},
\end{eqnarray}
where $\hat{S} = \vec{S}/S$. The first relation in Eq. 5 justifies the nomenclature, and the second relation emphasizes that the continuous set $\big\{|\hat{n}(\theta,\phi)\rangle\big\}$ constitutes an over-complete basis. The resolution of Identity (statement of completeness) takes the form
\begin{equation}
\frac{(2S+1)}{4\pi}\int d\Omega \vert \theta, \phi \rangle \langle \theta, \phi \vert \equiv I;
d\Omega = \sin \theta d\theta d\phi
\end{equation}
where $I$ is the identity operator in the Hilbert Space (of dimension $(2S+1)$ in this case).
Importantly, it follows from the second expression in Eq. 5 that the inner product between any two distinct states,
$\vert\langle\hat{n}|\hat{n'}\rangle \vert \rightarrow 0$ as $S \rightarrow \infty$, i.e., the states become mutually orthogonal in the limit of large spin. Since the manifold of SCS, represented by $\big\{\hat{n}(\theta,\phi)\big\}$, is identical to the phase space -- the two sphere ${\cal S}^2$ -- for a classical spin (apart from a scale factor), one says that the classical limit is attained in the large $S$ limit.

Being over-complete, when used as a basis, SCS allow any state to be completely expressed in terms of its diagonal elements
$
F(\hat{n})\equiv \langle\hat{n}|\rho|\hat{n}\rangle
$
\footnote{In contrast, if we were to employ a standard complete basis instead, one would have to specify the off diagonal elements as well.}. The functions $F(\hat{n})$ are the same as the well known $Q$ representation of the state in the coherent basis\cite{perel}.
Let us illustrate this for single qubit and two qubit states.
Considering the former, we have
\begin{equation}
\rho = \frac{1}{2}(I+ \vec{\sigma}\cdot \vec{P}) \longleftrightarrow F(\hat{n}) =\frac{1}{4\pi} (1 + \vec{P}\cdot \hat{n}),
\end{equation}
with the correspondence $\vec{\sigma} \longleftrightarrow \hat{n}$ making the bijective mapping between $\rho$ and $F$ clear. Explicitly, the vector $\vec{P}$ that characterizes the state is equivalently given by
\begin{equation}
\vec{P} = Tr\{\rho \vec{\sigma}\} = 3\int \hat{n}F(\hat{n})d\Omega.
\end{equation}

Similarly, a two qubit state given by
\begin{equation}
\rho^{AB} = \frac{1}{4} \big\{ I+ \vec{\sigma}^A\cdot \vec{P}_A + \vec{\sigma}^B\cdot \vec{P}_B +
\sigma^A_i\sigma^B_j \Pi_{AB}^{ij} \big\}
\end{equation}
has a one-one correspondence with its diagonal elements
\begin{eqnarray}
F(\hat{m},\hat{n}) & = & \frac{1}{4}\langle\hat{m}\otimes\hat{n}|\rho^{AB}|\hat{m}\otimes\hat{n}\rangle \nonumber \\
& \equiv & \frac{1}{(4\pi)^2}\big\{1+ \vec{P}_A\cdot \hat{m} + \vec{P}_B\cdot \hat{n} \nonumber \\
+& &\Pi_{AB}^{ij}\hat{m}_i\hat{n}_j\big\}.
\end{eqnarray}
which, again, clearly exhibits the bijective mapping between $\rho^{AB}$ and $F(\hat{m},\hat{n})$, defined
over the direct product space ${\cal S}^2 \otimes {\cal S}^2$.
The correspondence extends to any bipartite system of two spins $S_1,S_2$.

More generally, explicit expressions for the state in terms of $Q$ functions, as also for the expectation value of any operator, are not difficult to derive and are well known. We refer the reader to the beautiful discussion in \cite{perel} for details. Suffice to mention here that traces in Hilbert space get transformed into appropriate weighted averages in phase space, as illustrated in Eq. 8.

It may be noted that $F(\hat{n})$ for a spin $S$ state admits multipole moments upto a maximum rank $k=2S$. It may also be noted (see Eq. 5) that a classical pure state, $F(\hat{n}) = \delta(\hat{n} - \hat{n}_0)$ is approached asymptotically when $S \rightarrow \infty$.

\section{Formulation}
\subsection{An apparent paradox}
The stage is set to formulate the problem. We first pose an apparent paradox. Let $\rho^{AB}$ be the state of a bipartite system of two spins, and $F(\hat{m},\hat{n})$ its associated diagonal function in the coherent basis. Being non-negative everywhere,
it has the significance of a probability density in the classical phase space. Therefore, it always admits a resolution of the form
$
F(\hat{m},\hat{n})=\sum_{i}p_{i}f_{i}(\hat{m})g_{i}(\hat{n}).
$

Clearly, the probability density functions $f_{i}(\hat{m})$ and $g_{i}(\hat{n})$ can themselves be obtained as $Q$ representations of some parent quantum states $\rho_{i}^{A}$ and $\rho_{i}^{B}$, in their respective coherent basis.
This would, therefore, appear to yield
a separable resolution of $\rho^{AB} $ of the form given in Eq.1,
even if the state is non-separable, leading to an apparent paradox. This is depicted in Fig.1. The notion of classical simulability unravels this riddle.
\begin{figure}[H]
\includegraphics[scale=0.60]{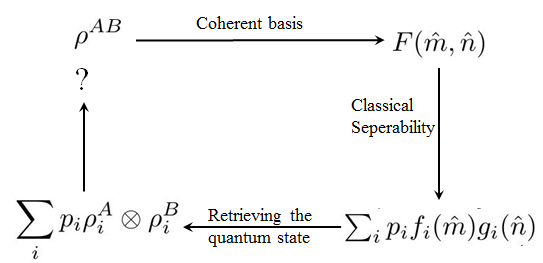}\caption{Schematic diagram indicating the apparent contradiction.}
\end{figure}

\section{Classical simulability}
To make precise what we mean by classical simulation, we introduce the concept of equivalence of states.

\subsection{ Equivalent States}
\noindent {\bf Definition 1}: Let $\rho \in {\cal H}^{M_1 M_2\cdots M_k}$ and $\rho^{\prime} \in {\cal H}^{N_1 N_2\cdots N_k}$ be two multipartite states with the same number of subsystems. We say that $\rho $ is equivalent to $\rho^{\prime}$, and write $\rho \cong \rho^{\prime}$, if both the states possess the {\it same} PDF as defined by the diagonal elements in their respective coherent bases.

As mentioned in Sect. IIB (see discussion following Eq. 7), it follows that the observables that respectively characterize the two states are also equivalent to each other. Operationally,
equivalent states reproduce the same values for all the observables, after suitable rescalings.
In other words, two equivalent states have the same information content. We illustrate this remark through some examples. We will be concerned only with bipartite systems here.

\noindent {\bf Example 1}: Consider spin systems which are completely
isotropic (unpolarized): $\rho^S = I/(2S+1)$. All of them possess the same PDF given by
$F(\hat{n}) = \frac{1}{4\pi}$, a uniform distribution over the unit sphere. Put differently, a classical distribution which is uniform over the sphere can correspond to a completely unpolarized spin system of any spin value (including $S=0$).

\noindent {\bf Example 2}: Consider the family of states ($S \neq 0$)
$\rho^S(\vec{P}) = \frac{1}{2S+1}\big(I + \hat{S}\cdot \vec{P} \big) \in {\cal H}^{2S+1}$. Recall that
$\hat{S}= \vec{S}/S$.
Positivity constraint implies that $\vert \vec{P}\vert \le 1$ for all $S$. These states
form a family of equivalent states, all yielding the same PDF, $F(\hat{n}) = \frac{1}{4\pi}(1+ \vec{P}\cdot \hat{n})$.

Note that this equivalence does {\it not} extend to properties such as the rank of the state. While a spin half state would be pure for the extremal value $\vert \vec{P} \vert =1$, it is mixed every where for all other spins.

\noindent {\bf Example 3}: The last example covers an important case of a bipartite system. Consider the family of correlated states for nonzero spins defined by
\begin{equation}
\rho^{S_1,S_2}[\alpha] = \frac{1}{ (2S_1+1)(2S_2+1)} \Big\{ I + \alpha \hat{S}_1\cdot \hat{S}_2 \Big\};~
-1 \le \alpha \le \frac{1}{3}.
\end{equation}
Clearly, all the states are equivalent to each other, yielding the ${\it same}$ PDF given by
\begin{equation}
F(\hat{m},\hat{n}) = \frac{1}{(4\pi)^2}\big\{ 1 + \alpha \hat{m}\cdot \hat{n} \big\}
\end{equation}

Note that the classical PDF is well defined over the larger range $|\alpha| \le 1$. The equivalence, however, does not extend to the full range since, in general, positivity of $\rho$ for any finite spin restricts the range to a smaller interval. For example, when $S_{1,2}= 1/2$, the upper bound on $\alpha$ is given by $1/3$. Thus a state $\rho^{S_1,S_2}[\alpha] \ncong \rho^{\frac{1}{2},\frac{1}{2}}[\alpha]$ when $\alpha > 1/3$. This will be used in our demonstrations in the next section.

\subsubsection{Extension to Observables} We mentioned that the notion of equivalence of states extends naturally to observables as well, with suitable re-normalizations or redefinition of ranges. Indeed, let
Let $O$ and $O^{\prime}$ be two observables acting on the Hilbert spaces ${\cal H}^{M_1 M_2\cdots M_k}$ and ${\cal H}^{N_1 N_2\cdots N_k}$ respectively. The two observables are equivalent if the equivalence $\rho \cong \rho^{\prime} \implies Tr\{\rho O\}
= Tr\{\rho^{\prime}O^{\prime}\}$.
We may extend the notation for equivalence to observables and write $O \cong O^{\prime}$.

As an illustration, consider the family of equivalent states $\rho^{S_1,S_2}[\alpha]$ defined in Eq 11. Let the observable $O=\sigma_z^A\otimes\sigma_z^A$ acting on ${\cal H}^{2,2}$. The observable $O^{\prime}$, acting on ${\cal H}^{2,2S+1}$ that is equivalent to $O$ is given by :
\begin{equation}
O^{\prime}=\frac{3S}{S+1}\sigma_z\otimes \hat{S_z}.
\end{equation}
Note that $O$ and $O^{\prime}$ are both rank 1 spin-spin correlations relevant to the two states $\rho = \rho^{\frac{1}{2}, \frac{1}{2}}(\alpha)$ and $\rho^{\prime}=\rho^{\frac{1}{2}, S}(\alpha)$. The rescaling mentioned is clearly reflected in the factor $3S/(S+1)$.

This demonstrates explicitly that states in a given equivalence class have the same information content.

\subsection{Classical Simulation}
\noindent {\bf Definition 2}: An state $\rho$ is said to admit classical simulation if it has an equivalent separable state $\rho^{\prime}$ \emph{which is finite dimensional}.

The restriction to finite dimensions is natural. Separable states in finite dimensions have a finite number of terms in their separable expansion \cite{cara}. This means that one needs a finite number of classical resources to simulate such a state.

Let a nonseparable state $\rho \cong \rho^{\prime}$, with $\rho^{\prime} = \sum_i p_i\rho^{\prime A}_i \times \rho^{\prime B}_i$. Clearly, in a generalized sense, $\rho$ is separable and the coefficients $p_i$ have the signiifcance of LHV. This accounts for the intriguing result of Werner that nonseparable states can admit a LHV interpretation \cite{werner}. We discuss this in greater detail in section VI. The following definition identifies states which do not admit an LHV description even in the generalized sense.

\vspace{0.1in}
\noindent {\bf Definition 3}: A state is exceptional if it does not admit classical simulation.

It follows from the examples discussed in the next section that violation of Bell inequality is necessary, but not sufficient, condition for a state to be exceptional.

\section{Three case studies}
Separable states admit classical simulation trivially and the interest is, therefore, in entangled states.
As mentioned, we examine simulability of three classes of states. Each of them spans the full range of entanglement, but differ in their coherence properties.
The results extend naturally to the full family of of each class related by local transformations.

\subsection{Isotropic states}
Isotropic two qubit states (Werner states) have the form given in Eq. 11 where both the constituents have spin $1/2$. Explicitly,
\begin{equation}
\rho^{W}[\alpha]=\frac{1}{4}\big\{I+\alpha\vec{\sigma}^{A}\cdot\vec{\sigma}^{B}\big\};~ -1 \le \alpha \le 1/3
\end{equation}
where the condition on $\alpha$ follows from positivity of the state. The states are mixed, except at $\alpha=-1$ where it is pure and fully entangled. Furthermore, they are also the so called $U$ invariant states in the sense of Werner. Recall that Werner was able to show that some $U$ invariant states may admit an LHV description \cite{werner}, even if they are entangled. It is therefore, natural that we ask if $\rho^W$ admit a classical simulation, and if so over what range. It follows from the PPT criterion that the state is entangled if $-1 \le \alpha <-1/3$ \cite{key-2,key-3}. This interval is naturally of interest to us.
The corresponding PDF $F^W[\alpha]$ is given by Eq. 11 which, of course, is separable everywhere. This means that there ought to exist higher dimensional bipartite states $\rho^{\prime}$which are (i) equivalent to a given $\rho^W[\alpha]$, and (ii) are separable. The restriction that the dimension be finite does not necessarily guarantee that classical simulation is possible for all values of $\alpha$.

\subsubsection{Construction of separable states equivalent to $\rho^W$}
Consider the family of isotropic bipartite states with $S \ge 1/2$,
\begin{equation}
\rho^{\frac{1}{2},S}[\alpha]=\frac{1}{2(2S+1)}\big\{I+\alpha\vec{\sigma} \cdot\hat{S}\big \}.
\end{equation}
Positivity of $\rho^{\frac{1}{2},S}[\alpha]~ \implies \alpha\in[-1,\frac{S}{S+1}]$, which contains the range for $\rho^W$ as its subset. Note also that as $S \rightarrow \infty$, $\alpha_{max} \rightarrow 1$. As demonstrated in section 3.1, $\rho^{\frac{1}{2},S}[\alpha] \cong \rho^W[\alpha]$ in the required range for $\alpha$.

That $\rho^W[\alpha]$ admits classical simulation everywhere except at $\alpha =-1$ is established by the following theorem:

\vspace{0.1in}
\noindent {\bf Theorem 1}: The isotropic states, $\rho^{\frac{1}{2},S}[\alpha]$ defined in Eq. 14
are separable in the range $|\alpha| \leq\frac{S}{S+1}$.

\vspace{0.05in}
\noindent {\bf Proof}: The proof is constructive, and exploits the isotropy property.
We start with the pure product state of maximum weight (in both the spins)
$\rho_z \equiv ( \vert \frac{1}{2} \rangle \otimes \vert S \rangle)( \langle \frac{1}{2} \vert \otimes \langle S \vert)$. In the irreducible tensor basis, it acquires the form
\begin{equation}
\rho_z=\frac{1}{2}\big\{I+\sigma_z\big\}\otimes\frac{1}{2S+1}\big\{I+\frac{3S}{S+1}\hat{S}_z+\sum_{k=2}^{2S}q_{z}^{k}S_{z}^{(k)}\big\}
\end{equation}
which emphasizes that the quantization is along the $z$ axis. The irreducible tensor operators are defined by $S^{(k)}_z = C_k (\vec{S}\cdot \vec{\nabla})^k r^k Y_{k0}(\hat{r})$. The normalization factors $C_k$ can be fixed conveniently, but fortunately it is not of much importance to us here.

Unlike $\rho^W$, $\rho_z$ admits polarizations of all ranks $k \le 2S$ and is anisotropic. We now construct similar states with the quantization axes along the $x$ and the $y$ directions and denote them by $\rho_{x,y}$ respectively.
The separable state $\bar{\rho} = \frac{1}{3} (\rho_x + \rho_y + \rho_z)$ obtained by their incoherent superposition has the form
\begin{eqnarray}
\bar{\rho} & \equiv & \frac{1}{3} (\rho_x + \rho_y + \rho_z) \nonumber \\
& = & \frac{1}{2(2S+1)}\big\{I+\frac{S}{S+1}\vec{\sigma}\cdot\hat{S}\big\} +\cdots
\end{eqnarray}
where the anisotropic terms, all of higher rank, are indicated by ellipsis. To eliminate the unwanted higher rank anisotropic terms, we perform a uniformization by averaging over the full sphere \footnote{ In practice, one needs only an averaging over a finite number of terms which depend on $S$, but that is not our prime concern here.}, which of course leaves the isotropic terms unchanged, and yields
\begin{equation}
\bar{\rho} \rightarrow \frac{1}{4\pi}\int \bar{\rho}d\Omega \equiv \rho^{\frac{1}{2},S}[\alpha_{max} =\frac{S}{S+1}]
\end{equation}
The required result follows from the observation that if $\rho^{\frac{1}{2},S}[\alpha]$ is separable then so is $\rho^{\frac{1}{2},S}[-\alpha]$. Thus the state $\rho^{\frac{1}{2}, S}[\alpha]$ is separable
over the range $ -\frac{S}{S+1} \le \alpha \le \frac{S}{S+1}$ for any spin $S$. Since PPT criterion is a necessary condition for separability, it follows that this state is non-separable over the complementary range $-1 \le \alpha < -\frac{S}{S+1}$
\footnote{ Interestingly in this case,
the PPT criterion has proved to be sufficient as well.}. The following theorem is thereby proved.

\noindent {\bf Theorem 2}: Only the Bell state is exceptional in the family of Werner states $\rho^W[\alpha]$. All other states admit a classical simulation. The minimum value of spin that simulates $\rho^W[\alpha]$ is the half integer nearest to $-\alpha/(1+\alpha)$. We simply write $S_{min} = -\alpha/(1+\alpha)$.

The entangling segment for $ \rho^{\frac{1}{2},S}[\alpha]$ shrinks to increasingly smaller intervals $\alpha \in [-1, -\frac{S}{S+1}) $ as S increases. Thus the degree of entanglement (non-separability) of Werner states, as represented by the corresponding $S_{min}$, is brought out vividly, with $S_{min} \rightarrow \infty$ rapidly as $\alpha \rightarrow -1$.
Fig 2 shows a segment of this behavior -- the devil's staircase -- in the range
$\alpha\in[-0.95,-0.85]$.

\begin{figure}[H]

\includegraphics[scale=0.35]{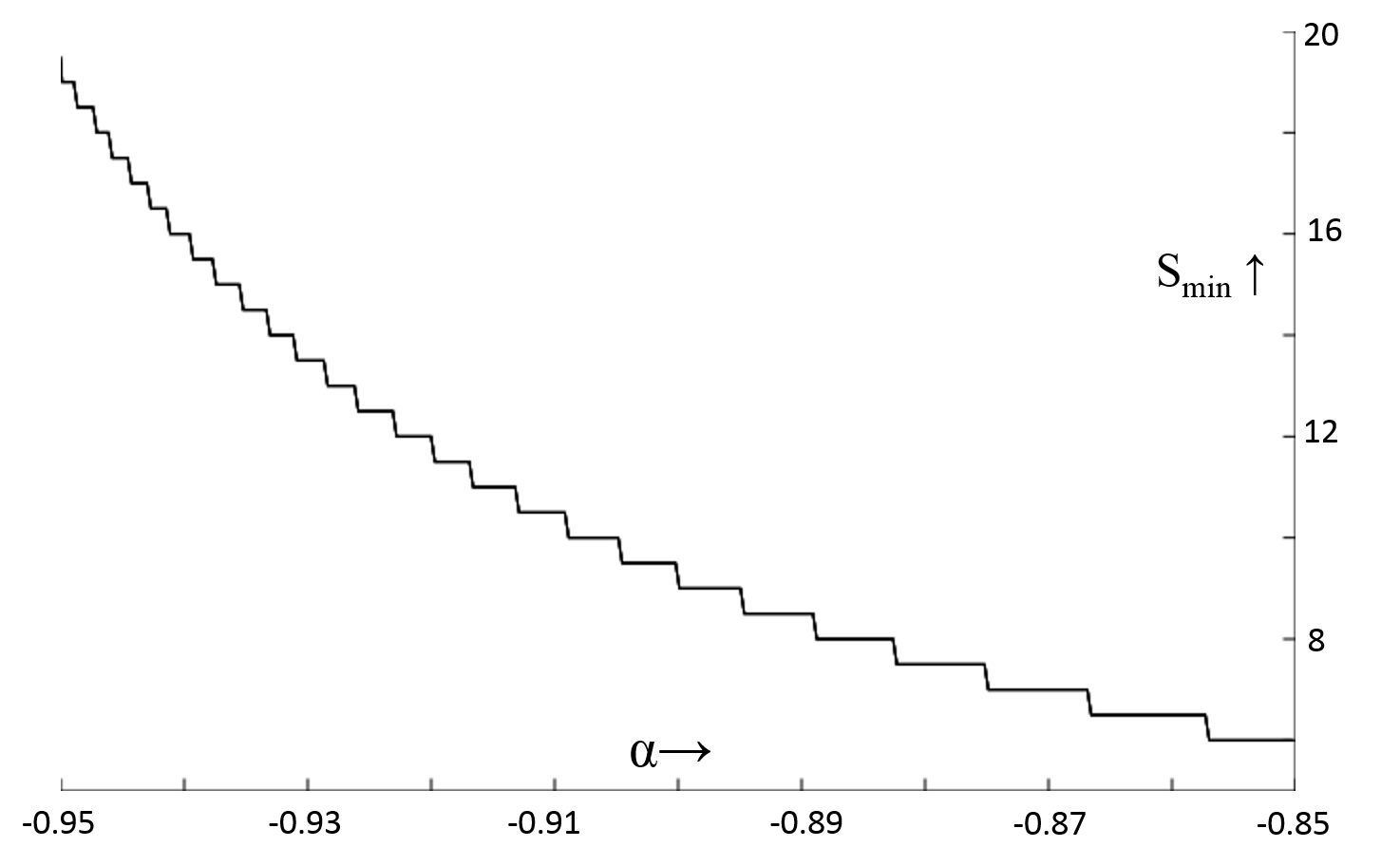}\caption{The higher spin classical equivalents of Werner states as a function of $\alpha$. The Bell State is at the pinnacle of this devil's staircase.}
\end{figure}

Theorem 2 has interesting implications since $\rho^W[\alpha]$ admits classical simulation at all $\alpha \neq -1$. Classical simulability in the region $\alpha \in [-1/\sqrt{2}, -1/3]$ explains the manner in which nonseparable states which respect Bell inequalities are local. But more surprisingly, the result that $\rho^W$ admits a classical simulation even when $\alpha < \alpha_0 = -1/\sqrt{2}$ throws up a surprise. This merits a careful discussion which we take up in Sect. VI, after discussing two more examples.

Before we discuss the next example, we note that this demonstration extends itself naturally in two directions. All mixed states which are related to $\rho^W$ by a local transformation also admit a classical simulation. Secondly, the demonstration establishes simulability of a larger class of states,
$ \rho^{\frac{1}{2},S}[\alpha]$. Simple though it seems, one may note that, even when they respect Bell inequality,  these higher dimensional states are not {\it not} $U-invariant$ as  the spins of the constituents are different. They are merely invariant under global $SU(2)$ transformations. Thus they are not covered by the example provided in \cite{werner}. Finally, it is easy to see that it is straight forward to generalize this result to states of the type $\rho^{S,S}[\alpha]$ and those obtained by operating local transformations on them.

\subsubsection{Resolution of the paradox}
The above example shows clearly how the concept of classical simulation provides a resolution of the paradox which was posed in Sect. IIIA. Consider again the mapping depicted in Fig. 1:
\begin{eqnarray}
\rho^{AB} & \rightarrow & F(\hat{m},\hat{n}) \nonumber \\
&= & \sum_i p_if_i(\hat{m})g_i(\hat{n}) \leftarrow \sum_ip_i \rho_i^A \otimes \rho_i^B. \nonumber
\end{eqnarray}
Separable states are self-equivalent. The states $\rho_i^A, \rho_i^B$ in the RHS of the equation above belong to the same Hilbert spaces as the constituents of the state $\rho^{AB}$. If the state is entangled, the dimensions of $\rho_i^{A,B}$ are higher, depending on the degree of entanglement. In the present example, with the exception of the Bell state, the dimensions for all states are finite implying that they are classical in the generalized sense. Only the Bell state is exceptional. We undertake a more detailed discussion of the connection between classical simulation and LHV in Sect. VI.

\subsection { Partially Entangled pure states}
Armed with the above results, we ask if partially entangled pure states also
admit classical simulation. If that were true, entanglement (in the sense of non-separability) would have the character of a universality. This question is of interest in view of the fact that they are known to violate Bell's inequality \cite{gisin,horo}. The following result dispels such a possibility. \\

\noindent {\bf Theorem 3}: All partially entangled pure states are exceptional.

\vspace{0.05in}
\noindent {\bf Proof}: We demonstrate that  every bipartite state of two equal spins $S$, which is equivalent
an entangled two qubit state, is non-separable.
It is convenient to employ the Schmidt basis and employ the canonical form
\begin{equation}
\vert \psi(\theta)\rangle = \cos\frac{\theta}{2}|\frac{1}{2} \rangle \otimes |\frac{1}{2} \rangle + \sin\frac{\theta}{2} |-\frac{1}{2} \rangle \otimes |-\frac{1}{2} \rangle.
\end{equation}
Its corresponding PDF is easily seen to be
\begin{eqnarray}
F_{\theta}(\hat{m},\hat{n}) &= & \Big\{1+\cos\theta(m_{z}+n_{z})+m_{z}n_{z} \nonumber \\
& & + \sin\theta(m_{x}n_{x}-m_{y}n_{y})\Big\}
\end{eqnarray}
The equivalent higher spin bipartite states, $\rho^{S,S} \cong \vert \psi (\theta)\rangle \langle \psi(\theta)\vert$, are necessarily mixed. They are easily found to be
\begin{eqnarray}
\rho^{S,S}(\theta) &=&\frac{1}{(2S+1)^{2}}\Big\{I+\cos\theta(\hat{S}_z^{A}\oplus \hat{S}_z^{B}) +
\hat{S}_z^{A}\otimes \hat{S}_z^{B} \nonumber \\
&& +\sin\theta(\hat{S}_{+}^{A}\otimes \hat{S}_{+}^{B}+\hat{S}_{-}^{A}\otimes \hat{S}_{-}^{B})\Big\}
\end{eqnarray}
where the normalized operators $\hat{S}_z=S_z/S;~~ \hat{S}_{\pm}=(S_x \pm iS_y)/(\sqrt{2}S)$.
Taking its partial transpose, we get
\begin{eqnarray}
\tilde{\rho}^{ S,S}(\theta) & = & \frac{1}{(2S+1)^{2}}\Big\{I+\cos\theta(\hat{S}_z^{A}\oplus \hat{S}_z^{B}) +\hat{S}_z^{A}\otimes \hat{S}_z^{B} \nonumber \\
& & +\sin\theta(\hat{S}_{+}^{A}\otimes \hat{S}_{-}^{B}+\hat{S}_{-}^{A}\otimes \hat{S}_{+}^{B})\Big\}.
\end{eqnarray}
If $\tilde{\rho}^{S,S}(\theta)$ were positive, so would $\Pi \tilde{\rho}^{S,S} \Pi$ be, where $\Pi$ is any projection operator. On the other hand, its projection onto
the subspace ${\cal H}_2$ spanned by the basis
${\cal B}= \big\{|S \rangle \otimes | -S \rangle; |S-1 \rangle \otimes | -S+1\rangle$\big\}, has the form
\begin{equation}
\tilde{ \rho}^{P}(\theta) =
\left(\begin{array}{cc}
0 & \sin\theta/S\\
\sin\theta/S & \frac{2S-1}{S^{2}}
\end{array}\right)
\end{equation}
which is nonpositive everywhere, except at
$\theta=0,\pi$, where the state is separable.
In short, the states $\rho^{S,S}(\theta) \cong \vert\psi(\theta)\rangle\langle \psi(\theta)\vert$ are not separable, however large the value of $S$ may be. Thus all partially entangled states are exceptional; this result is in conformity with violation of Bell inequality \cite{gisin}. Note however that the negative eigenvalue
$\lambda_- \rightarrow 0$ as $S \rightarrow \infty$.

Thus two states with the same entanglement can behave differently with respect to classical simulation. All Werner states with the exception of the Bell state are simulable. In contrast, all pure states with the exception of separable states are exceptional.

\subsection { A class of rank two states}
We have seen that all partially entangled isotropic states are mixed and may be simulated classically. It is also settled that partially entangled pure states cannot be simulated classically. To show that the interplay of entanglement and mixedness is more intricate than what the two examples may suggest, we construct a class of rank two mixed states which are again exceptional.

\vspace{0.1in}
\noindent{\bf Theorem 4}
The family of states
\begin{equation}
\rho(\mu; \theta, \theta^{\prime}) = \mu \vert \psi(\theta) \rangle \langle \psi(\theta)\vert +
(1-\mu) \vert \psi(\theta^{\prime}) \rangle \langle \psi(\theta^{\prime})\vert
\end{equation}
is exceptional almost everywhere in the parameter space.

\vspace{0.05in}
\noindent {\bf Proof}: The proof mimics the one employed for pure states. The construction of $\rho^{S,S} \cong \rho(\mu, \theta, \theta^{\prime})$ proceeds the same way as for the pure states.  After taking a partial transpose, the projected operator $\tilde{\rho}^P (\mu,\theta, \theta^{\prime})$ in the two dimensional subspace spanned by ${\cal B}$ defined in the previous subsection has the form
\begin{equation}
\tilde{\rho}^P (\mu,\theta,\theta^{\prime}) = \mu \tilde{\rho}^P(\theta) +(1-\mu) \tilde{\rho}^P(\theta^{\prime})
\end{equation}
which again, is nonpositive almost everywhere.


\section{Relation to Bell Inequality}
The three case studies bring out a number of interesting features of non-classicality, which merit a fuller comparison with the inferences drawn from violation of Bell inequalities.

\subsubsection{Werner States}
Consider first the isotropic Werner states, whose properties are summarized in theorem 2. There is, indeed, no conflict between classical simulation and locality when $\alpha \ge \alpha_0 = -1/\sqrt{2}$.
Our construction of equivalent separable states merely furnishes an explicit LHV description. It reconciles naturally the coexistence of nonseparability and locality which was discovered by Werner \cite{werner}. But for states which violate the inequality, we see that only the Bell state defies a classical description in our scheme, while all states in the interval $-1 \le \alpha < \alpha_0$ are deemed nonlocal. This difference merits a careful study, especially since the proof of Bell inequality is unassailable, and also because the demonstration here is explicitly constructive.

We believe that this can be traced, in part, to the difference in the definition of classicality/non-classicality in the two approaches. Recall that the formulation of standard LHV description in terms of inequalities \cite{bell,chsh} makes the crucial assumption
that the expectation values of single particle observables agrees both in the quantum state, as well as in the classical state that reproduces the correlations. Indeed, for the correlation considered for the singlet state in \cite{bell},
\begin{eqnarray}
P(\hat{a}, \hat{b})& \equiv & \langle \Psi_s \vert \vec{\sigma}_A \cdot \hat{a}~\vec{\sigma}_B \cdot \hat{b} \vert \Psi_s \rangle \nonumber \\
&= & \int d\lambda \rho({\lambda})A(\hat{a}, \lambda)B(\hat{b}, \lambda),
\end{eqnarray}
it is required that $|A(\vec{a}, \lambda)| \le 1$ and $|B(\vec{a}, \lambda)| \le 1$ since the Pauli matrices in the LHS have eigenvalues of unit norm. This restriction merely reflects the assumption that the constituents of the microstates that give rise to the correlation $P(\hat{a},\hat{b})$ are also essentially spin half systems with \emph{additional} hidden variables $\lambda$. Since locality, and not determinism is on focus here, we may further note that
the assignments $A(\vec{a}, \lambda) =\pm 1$ and $B(\vec{a}, \lambda) = \pm 1$ represents a deterministic scenario while allowing the full range of values over $[-1,+1]$ makes it non-deterministic.

Classical simulation defined here is not subject to  such a restriction. The notion of equivalence raises the question
if a given quantum (non-separable) state can be mimicked by a classical (separable) state, howsoever
large a spin it may possess. In particular, we do not require that the classical equivalent should comprise microstates belonging to equivalent qubits.

For this reason, the equivalent observables -- the counterparts of $A,B$ in Eq. 26 -- are {\it not} required to be bounded by unit norm. This has been illustrated explicitly in Eq. 13. We believe that this difference can be a contributing factor that can make states that violate Bell inequalities ($ \alpha < \alpha _0$) classically simulable. Indeed, at $\alpha_0$, the equivalent spin $S_{min} =5/2$ and it follows from Eq. 13 that the bound on observables is given by $15/7 >1$ which lends credence to the possibility.

As mentioned, it is not surprising that $\rho^W$ is simulable when $\alpha > \alpha_0$. For they may be separable, but they still respect Bell inequality. However, we do note that our construction does {\it not} give equivalent observables which are bounded by unit norm in the entangled local segment $\alpha_0 < \alpha < -1/3$. Presumably, there exist other more optimal constructions in this region which bound the equivalent observables by unit norm. The observables must necessarily belong to a higher spin since otherwise the state would be separable.

In short, the condition for classical simulability is weaker than the condition for locality as formulated in \cite{bell}. For that reason, the condition on quantumness becomes more stringent. Only
the fully entangled Bell state survives the criterion in this example.

\subsubsection{Partially entangled pure states and their rank 2 admixtures}
The next two examples further clarify, albeit in a qualitative manner, the role of coherence in determining whether a state can be simulated classically. It follows from Theorem 3 that a violation of Bell inequality is necessary and sufficient for a state to be exceptional. In the same vein, theorem 4 shows that exceptionality can survive even if we admit some decoherence, as exemplified by the class of rank 2 states obtained by taking an incoherent superposition of two pure states. Recall that Werner states, in contrast, have a minimum rank 3. The exact role played by coherence and entanglement in determining the conditions for a state to be exceptional needs further investigation.

\section{Quantum versus Classical resources}
The present approach allows us to estimate the resources that are required of a classical system to simulate a quantum system. Recall that by classical we mean quantum systems which are described by separable states. We treat simulable and exceptional states separately.

\subsection{Simulable states}
It is not difficult to bound the maximum number of hidden variables,$N_h$ that are required to characterize a state that admits a classical simulation. It is simply given by minimum number of
terms in the separable expansion of its equivalent state!. Indeed, if the equivalent separable state belongs to ${\cal H}^M \otimes {\cal H}^N$, then $N_h \le (MN)^2 + 1$ \cite{cara}. Thus, the number for non-separable $\rho^{W}[\alpha]$ is given by
$$N_h= 4(2S_{min} +1 )^2 +1 = 4(1-\alpha)^2/(1+\alpha)^2.
$$
If a state is infinitesimally
away from the Bell state, $\alpha = -1 +\epsilon$, the number of LHV grows as
$N_h(\epsilon) \sim 1/\epsilon^2$, showing the quadratically diverging number of classical resources
required to simulate the state.

\subsection{Exceptional States: approximate simulation}
Exceptional states cannot be simulated classically, however large the number of LHV may be.
Yet we may ask what the number of classical resources required to mimic an exceptional state upto an error $\epsilon$ would be.
\subsubsection{The Bell State}
It follows from Eq. 17 that if
a Werner state is infinitesimally away from the Bell state, $\alpha = -1 +\epsilon$, then $S_{min} \sim 1/\epsilon$, which can be interpreted as the dimension of the classical system that mimics a Bell state with an error $\epsilon$. Accordingly, the number of LHV that approximately describe the state also scales like $N_h(\epsilon) \sim 1/\epsilon^2$ signifying a quadratic divergence as the error $\epsilon \rightarrow 0$.

\subsubsection{Partially entangled pure states}
A similar analysis is possible for partially entangled pure states. It may be seen from Eq. 22 that as $S \rightarrow \infty$, the state approaches its classical limit. Indeed, consider the projected state shown in Eq. 22. As $S \rightarrow \infty$, its negative eigenvalue $\lambda_- =-\epsilon \rightarrow 0$, meaning that the state becomes more and more classical. Denoting the classical limit by $\rho_c$, we find that $\rho_c $ approximates $\rho^{SS}(\theta) \cong \rho(\theta)$ with an error $\epsilon$. The corresponding number of LHV are given given by
$$
N_h(\epsilon) \sim 1/ \epsilon_{eff}^2 = \sin^2 \theta/\epsilon^2.
$$
Note that the numerator is a measure of entanglement of the state. It is significant that this result holds even if the state is only infinitesmally away from a separable state, i.e., $\sin{\theta} \approx \theta \neq 0$.

\subsubsection{Rank two admixtures}
A straightforward adaptation to the rank two states (discussed in Sect. VC) $\rho(\mu,\theta,\theta^{\prime})$ yields
$$
N_h(\epsilon) \sim 1/ \epsilon_{eff} ^2 = \big\{\mu\sin \theta +(1-\mu) \sin \theta^{\prime}\big\}^2/\epsilon^2.
$$

\subsection{Quantum Dynamics and its classical simulation}
Finally, a brief digression on the possible implications of the results on quantum dynamics. In the language of quantum information, it translates into quantum circuits and measurements. There is a wealth of results in the literature on quantum computations which are intrinsically classical. Introducing the concept of matchgates, Valiant showed how a class of quantum computers could be efficiently simulated by a classical computer in polynomial time \cite{valiant,devin,knill}. These gates are known to be highly entangling. In a similar vein, Bartlett et. al. have obtained sufficient conditions for efficient classical simulation of quantum processes which involve continuous variables \cite{bartlett}. This result is an extension of the Gottessmann-Knill theorem \cite{gk} which states similar results for computation involving a restriction on the allowed gates. Parallel studies \cite{virmani} have dicussed circumstances -- such as losses, decoherence and fault tolerance -- under which a quantum computer becomes classical.

The same problem acquires a different complexion in the present framework.
Quantum dynamics which involves evolution of states is unitary. The question that naturally arises is the extent to which the unitary evolution can be simulated by a classical evolution, which would presumably be local in the following sense. If we approximate both the initial and final states by classical (separable) states with an error $\epsilon$, the unitary evolution would be equivalent to, within the same error margin, a "classical" evolution. The classical evolution would be "exact", though highly non-standard, if the states can be simulated classically. In any case, the evolution involves, in general, a transition between phase spaces of dimensions that could be vastly different. It may be expected that the number of resources required would yield an appropriate measure of complexity and efficiency.

\section{Conclusions}
In summary, classical simulation of quantum states --
-- seeking equivalent separable states in higher dimensions -- provides a new and a more rigorous way of characterizing quantumness of a state. It agrees with the Bell inequality criterion for pure states; however, It yields a weaker condition for mixed states to be local (simulable), and consequently a more stringent condition for them to be strictly non local (exceptional). It provides a natural explanation of Werner's result \cite{werner} that non-separable states can still be local. At this stage, our criterion is still is not completely operational -- since we do not possess a general method of verifying if a state is exceptional or not. Yet, it allows, in principle at least, of identifying the number of
LHV required to describe a classically simulable state, and also to mimic an exceptional state with an error $\epsilon \neq 0$. Finally, the present study points to the importance of the role of coherence, along with entanglement, for a state to be genuinely exceptional. This may not be without significance since in many a propsal, the notion of quantumness appears to be context dependent, as evidenced by the large number of proposals to describe entanglement and quantum correlations\cite{wootters, nielsen,zurek,horo2,usha1,shanth,plenio,horo3,augu}.
Given that there are so many perspectives to understand non-locality, quantum correlations and entanglement, one may safely conclude that the subject of classical quantum correspondence continues to remain a rich topic for further investigations and insights.

It is a pleasure to thank Professor A K Rajagopal for fruitful discussions.

\medskip{}


\begin{thebibliography}{References}
\bibitem{sch} E Schr\"odinger, \newblock\emph{Naturwissenschaften} 23, 807 (1935)
\bibitem{epr} A Einstein, B Podolsky and N Rosen, \newblock\emph{Phys. Rev} 47, 777 (1935)
\bibitem{bohm} D Bohm and Y Aharanov, \newblock\emph{Physical Review}, 108, 1070 (1957)
\bibitem{bell} J S Bell, \newblock\emph{Physics} 1, 195 (1964)
\bibitem{chsh} J F Clauser, M A Horne, A Shimony and R A Holt, \newblock\emph{Phys. Rev. Lett.} 23, 880 (1969)
\bibitem{ch} J F Clauser and A Shimony, \newblock\emph{Rep. Prog. Phys.} 41, 1881 (1978)
\bibitem{werner} R. F. Werner, \newblock\emph{Phys. Rev. A} 40, 4277
(1989).



\bibitem{sudarshan} J R Klauder and E C G Sudarshan, \newblock\emph{Fundamentals of Quantum Optics}, (W A Benjamin, New York, 1968) Chapter 7, p105

\bibitem{arechhi} F T Arechhi, E T Courtens and Robert Gilmore, \newblock\emph{Phys. Rev. A} 6, 2211 (1972)

\bibitem{perel} A. Perelomov, \newblock\emph{Generalized coherent states and their applications}, (Springer Verlag, Berlin, 1986) Chapter 4, p54. For more specialized reviews see \cite{arechhi,zhang}


\bibitem{zhang} J R Klauder and G Skagerstam, \newblock\emph{Coherent States, World Scientific, Singapore}, 1985; Wei-Min Zhang, Da Hsuan Weng and Robert Gilmore, \newblock\emph{ Revs. Mod. Phys.} 62, 869 (1990). For a different perspective on SCS, see

\bibitem{scully} M O Scully and W\'{o}dkiweicz, \newblock\emph{Found. Phys}, 24, 85 (1994)

\bibitem{cara} This follows from the celebrated Caratheodory theorem on convex hulls: C Caratheodory, \newblock\emph{Rendoconti del Circolo Matematico di Palermo}, 32, 193 (1911). See also G Dahl, \newblock\emph{Introduction to Convexity}, University of Oslo, Unpublished (2009) for a pedagogic exposition.


\bibitem{key-2}A. Peres, \newblock\emph{Phys. Rev. Lett}. 77:1413
(1996)

\bibitem{key-3}M. Horodecki, P. Horodecki, R. Horodecki, \newblock\emph{Physics
Letters A} 223, 1 (1996)

\bibitem{gisin} N Gisin, \newblock\emph{Phys. Lett. A} 154, 201 (1991)

\bibitem{horo} Horodecki et al., \newblock\emph{Phys. Lett. A} 200, 340 (1995)

\bibitem{valiant} L. Valiant, \newblock \emph{in Proceedings of the ACM Symposium on the
Theory of Computing} ~ACM Press, New York, 2001, p. 114;
see http://www.deas.harvard.edu/~valiant. For a physical realization in terms of noninteracting fermions, see
\bibitem{devin} Barbara. M. Terhal and David. P. Divincenzo, \newblock\emph{Phys. Rev. A} 65, 032325 (2002) and for its relation to fermionic linear optics, see

\bibitem{knill} E Knill {\it unpublished} \newblock\emph{arXiv: quant-ph/0108033} 2001


\bibitem{bartlett} S. D. Bartlett, B. C. Sanders, S. L. Braunstein and K. Nemoto, \newblock\emph{Phys. Rev. Lett. } 88, 097904 (2002)

\bibitem{gk} D Gottesman, \newblock\emph{ArXiv: quant-ph/9807006; Proc. Int. Coll. on roup theoretical methods in Physics, Eds. S P Corney et. al}, (Cambridge, MA, International Press), 32 (1999)



\bibitem{virmani} S. Virmani, S. F. Huelga, and M. B. Plenio, \newblock\emph{Phys. Rev.
A,}71, 042328 (2005); N. Ratanje and S. Virmani, \newblock\emph{Phys. Rev. A} 83, 032309 (2011)


\bibitem{wootters}W. K. Wootters, \newblock\emph{Phys. Rev. Lett}.
80, 2245 (1998),

\bibitem{nielsen} M A Nielsen, \newblock\emph{Phys. Rev. Lett} 83, 436 (1999)

\bibitem{horo2} M Horodecki, \newblock\emph{Quant.Information.Comp} 1, 3 (2001)

\bibitem{zurek} H. Ollivier and W. H. Zurek, \newblock\emph{Phys. Rev. Lett}. 88, 017901, (2001)

\bibitem{plenio} Martin B Plenio and S Virmani, \newblock\emph{Quant.Inf. Comp} 7, 1 (2007);

\bibitem{shanth}Shanthanu Bhardwaj and V Ravishankar, \newblock\emph{Phys. Rev. A} 77, 022322 (2008)

\bibitem{horo3} R Horodecki, M Horodecki, P Horodecki and K Horodecki, \newblock\emph{Rev. Mod. Phys.} 81, 865 (2009)

\bibitem{augu} R Augusiak et.al, \newblock\emph{Phys. Rev. Lett.} 107,070401 (2011)

\bibitem{usha1} A R Usha Devi and A K Rajagopal, \newblock\emph{Phys. Rev. Lett} 100, 140502 (2008);
A R Ushadevi et. al., \newblock\emph{Quantum Inf Process} 12, 3717 (2013)



\end{thebibliography}
\end{document}